# Noncommutative Simple Harmonic Oscillator in Quantum Gravitational Well

Farhad Zekavat[1]

Islamic Azad University, Central Tehran Branch

**Abstract:** This work is mainly based on some theoretical surveys on two dimensional quantum gravitational well, considering harmonic oscillator potential causes an effective plank constant. We find that there is a similarity between two different cases, potential free and oscillator potential in noncommutative phase space in commutative momentum, but in noncommutative, non zero space and momentum The Hamiltonian would a little bit different in energy levels.

## Introduction

Noncommutativity have some advantages in theory for both quantum mechanics and quantum fields. To solve some infinities in both contexts we can use a something like $\theta_{ij}$ which is a result of violation of the Lorentz symmetry in commutative of $x^\mu$:

$$[x^\mu, x^\nu] = i\theta^{\mu\nu} \qquad (1)$$

Which $\theta^{\mu\nu}$ is antisymmetric. For phase space transformation and 4-dimentional space-time

$$[X^\mu, P^\nu] = i\hbar\delta^{\mu\nu}$$

$$[P^\mu, P^\nu] = i\eta^{\mu\nu} \qquad (2)$$

$\eta^{\mu\nu}$ is also antisymmetric [9]. Also we can easily back to ordinary commutative algebra. $\theta$ and $\eta$ are momentum and space coordinates in noncommutative geometry.

This work is based on 2-dimentional space-time. There were some papers which proposed 2d studies of momentum phase of gravitational quantum well and main idea is the based on [2, 3].

We have considered phase space and an additional physical term or oscillator to this quantum mechanical system to see what would be the results and how these assumptions affect a physical system. In section I, is a review of basics and previous works. In section II, we insert kinetic energy of simple harmonic oscillator (SHO) and compare the results for effective plank constant.

In section III, we insert both NC space and momentum and also previous section assumption of presenting kinetic energy of SHO. Then in section IV, the $\hbar_{eff}$ for this noncommutative quantum gravitational well (NCQGW).

## I. Noncommutative quantum mechanics

For a 2 dimensional quantum mechanical system NC:

[1] f.zekavat@iauctb.ac.ir

$$[x, y] = i\theta$$

$$[p_x, p_y] = i\eta \quad (3)$$

$$[x_i, p_j] = i\hbar \delta_{ij}$$

which $i=x$ and $j=y$ here. These relations can be obtained from linear transformations:

$$x = x' - \frac{\theta}{2\hbar} p'_y$$

$$y = y' + \frac{\theta}{2\hbar} p'_x \quad (4)$$

$$p_x = p'_x + \frac{\eta}{2\hbar} y'$$

$$p_y = p'_y - \frac{\eta}{2\hbar} x'$$

If we apply equations (4) to (3), we may find:

$$[x_i, p_j] = i\hbar \left(1 + \frac{\theta\eta}{4\hbar^2}\right) \delta_{ij} \quad (5)$$

Comparing two commutation relations, we find that there is a $\hbar_{eff}$ which depends a the noncommutative parameters $\theta$ and $\eta$. we can easily back to ordinary phase by giving $\theta$ and $\eta$, zero.

$$\hbar_{eff} = \hbar(1 + \xi) \quad (6)$$

where $\xi = \frac{\theta\eta}{4\hbar^2}$. Therefore, if $\xi \ll 1$, we live in ordinary quantum mechanics. So NC effect is in second order. If we rewrite equation (4):

$$X' = C\left(x + \frac{\theta}{2\hbar} P_y\right), \quad y' = C\left(y - \frac{\theta}{2\hbar} P_x\right), \quad P'_X = C\left(P_X - \frac{\eta}{2\hbar} y\right), \quad P'_y = C\left(P_y + \frac{\eta}{2\hbar} x\right) \quad (7)$$

which $C = (1 - \xi)^{-1}$. First of all, we start with Hamiltonian accompanied with gravitational force,

$$H' = \frac{p'^2_x}{2m} + \frac{p'^2_y}{2m} + mgx' . \quad (8)$$

One can find that horizontal motion in experiment [4, 5, 6] and theoretical [12] in 2 dimensions. Applying equation (7) to (8),

$$H = \frac{C^2 p^2_x}{2m} + \frac{C^2 p^2_y}{2m} + \frac{C^2 \eta}{2m\hbar}(xp_y - yp_x) + \frac{c^2}{8m\hbar^2}\eta^2(x^2 + y^2) + mgC\frac{\theta p_y}{2\hbar} + mgCx \quad (9)$$

We can substitute a better term for $\bar{p}_x$ and $p_y$ terms by defining

$$\bar{p}_y \equiv Cp_y + \frac{m^2 g\theta}{2\hbar} , \quad \bar{p} \equiv Cp_x . \quad (10)$$

This noncommutative Hamilton can be written as:

$$H = \frac{\bar{p}^2_x}{2m} + \frac{\bar{p}^2_y}{2m} + \frac{C\eta}{2m\hbar}(x\bar{p}_y - y\bar{p}_x) + \frac{C^2 \eta^2}{8m\hbar^2}(x^2 + y^2) + mgCx - mgx\frac{\theta\eta}{4\hbar^2}x . \quad (11)$$

Two last terms belong to gravitational potential

$$mgCx - mgC\frac{\theta\eta}{4\hbar^2}x = mgC(1 - \zeta)x = mgx \quad (12)$$

There two terms are noncommutative effects corresponds two commutative algebra.

## II. SHO and oscillations

In a quantum mechanical system we may have expected kinetic energy, especially in bound states.

$$H = \frac{P_x{}^2}{2m} + \frac{P_y{}^2}{2m} + \frac{k}{2}(x^2 + y^2) \tag{13}$$

Here we start to consider another additional and expected term in a quantum system. The oscillation potential part (typical $\frac{kx^2}{2}$)

$$H' = \frac{p'_x{}^2}{2m} + \frac{p'_y{}^2}{2m} + mgx' + \frac{kx'^2}{2} + \frac{ky'^2}{2} \tag{14}$$

Note that have we assume

$$p'_x = p_x \quad , \quad p'_y = p_y \tag{15}$$

Means noncommutative and commutative momentum are equivalent or simply we are in ordinary phase space for p, but still we use equation (4) to proceed. Using space to transformations we have,

$$H = \frac{p_x^2}{2m}\left(1 + \frac{C^2 k\theta^2}{4\hbar^2}\right) + \frac{p_y^2}{2m}\left(1 + \frac{C^2 k\theta^2}{4\hbar^2}\right) + \frac{kC^2}{2}(x^2 + y^2)$$

$$+ mgCx + \frac{kC^2\theta}{2\hbar}(xp_y - yp_x) + mgC\left(\frac{\theta p_y}{2\hbar}\right) \tag{16}$$

To write a clear form of Hamiltonian we define

$$\bar{p}_y = C'p_y + \frac{(m^2 g\theta)}{2\hbar},$$

$$\bar{p}_x = Ap_x = c(1 + \frac{k\theta^2}{4\hbar^2})p_x \tag{17}$$

Therefore

$$H = \frac{\bar{p}_x^2}{2m} + \frac{\bar{p}_y^2}{2m} + mgCx + (coeffi)(x\bar{p}_y - y\bar{p}_x) + \frac{k}{2}(x^2 + y^2) \tag{18}$$

As it sounds here there is a difference between equations (11) and (14). From the equation (11) we find that by applying noncommutative algebra to our considered quantum mechanical system (eq.(8)), Hamiltonian would behave like a quantum system which have oscillator potential (14). But if we have a quantum mechanical system with oscillator potential in commutative context, and convert commutative to noncommutative algebra, the result would be the equation (18) and we find that there would be no difference between (11) and (18), by means of the Hamiltonian of these two different (with and without) Oscillator potential behave the same in noncommutative geometry. The only difference between (11) and (18) is Potential energy in (11) comes from noncommutative algebra, but in (18) it is the potential energy which belongs to our commutative (and supposed physical) from the beginning. Noncommutative effect here shows the physical behavior in angular momentum term that we did not find in our ordinary quantum system. Thus a NC magnifying glass reveals a $l_z$ -like property in both free particle and oscillator potential.

## III. Non-commutative space- momentum: the general case

If we imply our quantum mechanical system in (14) both space and momentum noncommutative transformations [ref.8 is helpful for all cases of NC parameters in deferent approaches and problems, specially on Landau problem] in (7) to find what would come out,

$$H = C^2\left(1 + \frac{km\theta^2}{4\hbar^2}\right)\frac{p_x^2}{2m} + C^2\left(1 + \frac{km\theta^2}{4\hbar^2}\right)\frac{p_y^2}{2m} + mgCx + mgC\left(\frac{\theta p_y}{2\hbar}\right) + \left(\frac{kC^2}{2} + \frac{\eta^2}{8\hbar^2}\right)(x^2 + y^2)$$

$$+ C^2\left(\frac{\eta}{2m\hbar} - \frac{k\theta}{2\hbar}\right)(xp_y - yp_x) \quad (19)$$

Or in shorthand form

$$H = C'\frac{p_x^2}{2m} + C'\frac{p_y^2}{2m} + mgCx + mgc\left(\frac{\theta p_y}{2\hbar}\right) + D(x^2 + y^2) + E(xp_y - yp_x) \quad (20)$$

Where

$$C' = C^2\left(1 + \frac{km\theta^2}{4\hbar^2}\right), \quad E = C^2\left(\frac{\eta}{2m\hbar} - \frac{k\theta}{2\hbar}\right), \quad D = \frac{kC^2}{2} + \frac{\eta^2 C^2}{8m\hbar^2} \quad (21)$$

Comparing to kinetic energy coefficients of (14) and (19)

$$(1-\xi)^{-1} = C' = C\left(1 + \frac{mk\theta}{4\hbar^2}\right)^{\frac{1}{2}} = \left(1 - \frac{\eta\theta}{4\hbar^2}\right)\left(1 + \frac{km\theta^2}{4\hbar^2}\right)^{\frac{1}{2}} \quad (22)$$

Easily by expanding the second binomial

$$\left(1 + \frac{mk\theta^2}{8\hbar}\right)^{\frac{1}{2}} \approx \left(1 + \frac{mk\theta^2}{8\hbar^2}\right) \quad (23)$$

Omitting further terms, we final that

$$\left(1 - \frac{\eta\theta}{4\hbar^2}\right)\left(1 + \frac{mk\theta^2}{8\hbar^2}\right) = 1 + \frac{mk\theta^2}{8\hbar^2} - \frac{\eta\theta}{4\hbar^2} - \frac{mk\theta^3\eta}{32\hbar^4} - (\ldots) \quad (24)$$

Again we omit the last term for $\hbar^{-4}$, so we have a clean expression and we find that it is physical. Therefore, taking

$$(1-\xi')^{-1} = 1 + \xi',$$

which we had it before (6),

$$\xi' = \frac{mk\theta^2}{8\hbar^2} - \frac{\eta\theta}{4\hbar^2} \quad (25)$$

Then

$$\hbar_{eff} = \hbar(1 + \xi) = 1 + \frac{mk\theta^2}{8\hbar^2} - \frac{\eta\theta}{4\hbar^2} \quad (26)$$

Compare (26) To [11], $\xi = \frac{\eta\theta}{4\hbar^2}$ and experimental parameter $\zeta_n = \left[\frac{3\pi}{z}\left(n - \frac{1}{4}\right)\right]^{\frac{2}{3}}$ [7].

We write again the Hamiltonian using (26),

$$H = \left(1 + \frac{mw^2}{8\hbar^2} - \frac{\eta\theta}{8\hbar^2}\right)(p_x^2 + p_y^2) + mg\left[\left(1 - \frac{\eta\theta}{4\hbar^2}\right) + \frac{\theta p_y}{2\hbar x}\right] + \left(\frac{mw^2}{2} - \frac{\theta^2}{8m\hbar^4}\right)(x^2 + y^2)$$

$$+(\frac{\eta}{2m\hbar} - \frac{mw^2\theta}{2\hbar})(x\bar{p}_y - y\bar{p}_x) \qquad (27)$$

The last term express $L_z$ as mentioned. As we can see, if we use $\hbar_{eff}$ in equation (26) in Hamiltonian, then it appears that someone is interrupting in our physical system! In our measurements will change a little bit by this interrupting cause.

For first order perturbation we have

$$V_1 = (\frac{\eta}{2m\hbar} - \frac{mw^2\theta}{2\hbar})xp_y \qquad (28)$$

It is equivalent to $\boldsymbol{B} = \hat{e}_z B$ that is perpendicular to this plane, in this case as people did in experiment. So

$$\frac{\eta}{2m\hbar} - \frac{mw^2\theta}{2\hbar} = \frac{q\boldsymbol{B}}{m} \qquad (29)$$

or squared NC momentum is

$$\eta = m^2 w^2 \theta + 2q\boldsymbol{B}\hbar \qquad (30)$$

Then we have frequency of the particle for the first order perturbation:

$$\omega = \frac{\sqrt{\eta/\theta - 2q\boldsymbol{B}\hbar/\theta}}{m} \qquad (31)$$

or the energy part of this part of frequency of spectrum will be:

$$E = \hbar\omega = \hbar\frac{\sqrt{\eta/\theta - 2q\boldsymbol{B}\hbar/\theta}}{m} \qquad (32)$$

Thus, in ordinary quantum mechanics, there would be ordinary frequency for particle. If the particle has minus charge such as electron, q= -e, and if it is a positive charged particle, the direction of inclination would be in the opposite direction (-B).

For neutral particles

$$\omega = \frac{\sqrt{\eta/\theta}}{m} \qquad (33)$$

or the energy part is:

$$E = \hbar\omega = \frac{\sqrt{\eta/\theta}}{m} \qquad (34)$$

Thus a minimal length is required or in non-minimal length, we may think of renormalization as expected, but regularization is the other way. A nice argument and calculation can be found in A. Kempf et al [5] and also M. Maggiore [11]. Also as original hints should be checked Pauli and Villars discussed and introduced about invariant regularization [16]. Dirac and Heisenberg hint brilliantly about what people are doing latter till now.

And for the second order of perturbation, regarding the direction of the particle, we have

$$V_2 = \frac{mw^2\theta^2}{8\hbar^2}P_x^2 - \frac{\theta\eta}{8\hbar^2}P_x^2 + \frac{\eta^2}{8m\hbar^2}x^2 \qquad (35)$$

First and second terms refer harmonics and the last term refers to gravity. It means the gravitational frequency is

$$\omega = \frac{|\eta|}{2m\hbar} \qquad (36)$$

Or for squared momentum in NC quantum mechanics:

$$|\eta| = 2m\hbar\omega \tag{37}$$

As it looks like for neutral particles, gravitational influence in the second order perturbation is

$$E_g = \hbar\omega = \frac{|\eta|}{2m} \tag{38}$$

For commutative case we have ordinary term (η=0), and $iy$ we do not have magnetic force included, there is no effect and particle from gravitation frequency an in ordinary phase space we see. But theoretically, as we see further Non-commutative quantum mechanics plays the rule of magnetic field or in other words, it is magnifying – glass to see a more real value which are compatible with experiments in presence of B-field.

For a fruitful discussion on non-commutative algebra and magnetic fields reference [8] would be enlightening, and for NC parameters [8 and 14].

**IV. results**

From the given plank constant in section III, and applying on the quantum system, on finds between initial and final states there would be a distortion or another measurement, like spying on the system. $V_1$ and $V_2$ reveal the nature of this distortion. Missing information is another meaning and discussed in Ref. [15] for a non-gravitational harmonic oscillator. I surveyed and found similar result for NC momentum eq.(29-30).

We have gravity instead of the Ref. [15] case but comparing to results the gravity has nothing to do with $V_1$. Also NC space was not enough to affect on this SHO, but momentum changes the result (section III).

In the presence of magnetic field, charged particles' frequency would be eq.(31). If there would be no external magnetic field, naturally NC space and momentum are responsible for the frequency as seen in eq. (33), and in some approaches intrinsic magnetic field for momentum noncommutativity appears. There is an important point that for SHO, there would be a reason to oscillations which a matter of many cases in the quantum physics and microscopic material surveys.

Here we may need a minimal length to contour infinity of frequency or explore fragmentation discussed in section III. In $V_2$, zero NC momentum can cause no frequency due to gravity; which means in canonical case or even no magnetic strength, we cannot measure gravitational effect (eq.36). Indeed, if there would be no NC momentum, we cannot measure $V_1$ frequency neither and it will be the canonical case in ordinary quantum mechanics.

**V. Outlook**

This work which is following the method of Ref [13] by Bertolami et al, is a short and fast study of two possibility: momentum transformation and its effect on spectrum and effective plank constant; also simple harmonic oscillator by the Ref [16], Pauli and Villars method is perturbative clearly. This two dimensional quantum system influenced by gravity which people did in some works before. There are some non-gravitational modes explored [11,12].

Relations between NC parameters has been discussed in Ref [13]. Also a related calculation has done by J. B. Geloun et al [9] in a background magnetic field, which is another non-gravitational mode, but similar to two dimensional harmonic oscillator.

This work will follow by analysis of perturbation and a theoretical discussion about NC parameters, magnetic fields and different point of views in the next work of mine.